\documentclass[preprint,superscriptaddress,nofootinbib,amsmath,amssymb,aps,longbibliography] {revtex4-1}
\usepackage{graphicx}
\usepackage{amsmath}
\usepackage{amsfonts}
\usepackage{mathtools}
\usepackage{color}
\usepackage{amssymb}
\usepackage{dcolumn}
\usepackage{bm}
\usepackage{microtype} 
\usepackage[linktoc=all]{hyperref}
\usepackage[capitalize]{cleveref}

\numberwithin{equation}{section}
\newcommand{\lag}{\mathcal{L}}
\newcommand{\ud}{\mathrm{d}}

\newcommand{\rb}{\mathrm{b}}
\newcommand{\mn}{{\mu\nu}}

\makeatletter
\DeclareRobustCommand{\rcite}[1]{%
  \rcite@aux#1,\@nil{#1}%
}
\def\rcite@aux#1,#2\@nil#3{%
  \if\relax#2\relax
    Ref.~\cite{#3}%
  \else
    Refs.~\cite{#3}%
  \fi
}
\makeatother

\begin{document}
	\title{Solitons in generalized galileon theories}
	
	\author{Mariana Carrillo Gonz\'alez} 
	\email{cmariana@sas.upenn.edu}
		\affiliation{Center for Particle Cosmology, Department of Physics and Astronomy,
		University of Pennsylvania, Philadelphia, Pennsylvania 19104, USA}
	\author{Ali Masoumi}
		\email{ali.masoumi@tufts.edu}
	\affiliation{Institute of Cosmology, Department of Physics and Astronomy,
		Tufts University, Medford, MA 02155, USA}
	\author{Adam R. Solomon}
		\email{adamsol@physics.upenn.edu}
			\affiliation{Center for Particle Cosmology, Department of Physics and Astronomy,
		University of Pennsylvania, Philadelphia, Pennsylvania 19104, USA}
	\author{Mark Trodden}
		\email{trodden@physics.upenn.edu}
			\affiliation{Center for Particle Cosmology, Department of Physics and Astronomy,
		University of Pennsylvania, Philadelphia, Pennsylvania 19104, USA}
	
	\date{\today}
	
	\begin{abstract}

We consider the existence and stability of solitons in generalized galileons, scalar field theories with higher-derivative interactions but second-order equations of motion. It has previously been proven that no stable, static solitons exist in a single galileon theory using an argument invoking the existence of zero modes for the perturbations. Here we analyze the applicability of this argument to generalized galileons and discuss how this may be avoided by having potential terms in the energy functional for the perturbations, or by including time dependence. Given the presence of potential terms in the Lagrangian for the perturbations, we find that stable, static solitons are not ruled out in conformal and (A)dS galileons. For the case of DBI and conformal galileons, we find that solitonic solutions moving at the speed of light exist, the former being stable and the latter unstable if the background soliton satisfies a certain condition.
	\end{abstract}
	
	\maketitle
	
	\tableofcontents
\section{Introduction}
Solitons were first observed in a hydrodynamical context \cite{Russell:1845,Korteweg:1895} nearly two centuries ago and have since been found in fields as diverse as condensed matter, cosmology, and particle physics. Solitons have been particularly useful for modeling hadrons due to their non-perturbative nature \cite{Friedberg:1976eg,Zahed:1986qz}. In condensed-matter systems, a variety of types of soliton have been observed in experiments \cite{PhysRevLett.94.184503,cmsol}. Solitonic solutions that arise in field theories can be generated in the early cosmos during a symmetry-breaking phase phase transition, and hence may play an important role in the evolution of our universe. Similarly, their existence can be highly constrained by cosmological observations \cite{Vilenkin:2000jqa, Nambu:1977ag}, and by comparing their production rate in the early universe with the present bounds on their abundance, it is possible (as long as inflation does not last too long) to constrain the underlying field-theoretic model.

In this paper, ``solitons" will refer specifically to non-trivial field configurations with finite energy that are localized in space and which do not dissipate over time. These classical field solutions typically have energies proportional to the inverse of the field's coupling constant, indicating their non-perturbative nature. Their existence is due to non-linearities in the field; they are not sustained by external sources. Some solitons, known as topological solitons, are stable due to a topological conserved charge. Another class consists of non-topological solitons \cite{Lee:1991ax}, such as Q-balls \cite{Coleman:1985ki}, whose existence is due to a conserved Noether charge. Non-topological solitons are stable because their energy is lower than any other configuration (including a collection of free particles) with the same charge. Also of interest are oscillons \cite{Bogolyubsky:1976yu, Graham:2006vy}--- metastable field configurations similar to domain walls or bubbles which are long-lived by virtue of oscillating in a specific mode. These are approximate solutions that slowly radiate away their energy.

The existence of topological solitons depends on the structure of the vacuum manifold. If the vacuum has sufficiently complicated topology, as measured by the nontriviality of certain homotopy groups, then solitons exist, and their stability is guaranteed since the boundary conditions for the soliton are topologically different from that of the physical vacuum state. Solitons may also arise in the presence of higher-derivative terms in an effective field theory; these can also lead to non-linearities stabilizing the soliton. Perhaps the best known example of this is in the case of skyrmions \cite{Skyrme:1961vq, Zahed:1986qz}.

To date, defects in the presence of non-canonical kinetic terms have been studied in only a handful of cases. In scalar field theories with non-canonical kinetic terms topological defects are called $k$-defects \cite{Babichev:2006cy,Adam:2007ij}. Here the non-triviality of the homotopy groups of the vacuum manifold is not enough to guarantee the existence of solitonic solutions. In some cases $k$-defects can mimic the field profile and energy density of defects in the corresponding canonical scalar field theory \cite{Andrews:2010eh}. Stable domain wall solutions were also found in particular higher-derivative scalar theories in \rcite{Bazeia:2014dva}.

With the notable exception of the Skyrme model, the aforementioned theories possess non-canonical derivative terms \emph{in addition to} a symmetry-breaking potential, so that the solitons are not sustained by non-canonical or higher-derivative terms alone. To construct new examples of solitons using solely derivative interactions, promising candidates are the \emph{galileon} theories, a class of scalar theories with higher-derivative interactions that lead to only second-order equations of motion \cite{Nicolis:2008in}.\footnote{Galileon theories exhibit a panoply of interesting effects. For example, their effects on nonlinear scales are screened due to the Vainshtein mechanism \cite{Babichev:2013usa}, they admit stable solutions violating the the null-energy condition \cite{Creminelli:2010ba,Hinterbichler:2012yn}, their cosmological solutions can self-accelerate, providing an alternative to dark energy \cite{Silva:2009km,Chow:2009fm}, and they do not get renormalized at any order in perturbation theory \cite{Luty:2003vm,Hinterbichler:2010xn,Goon:2016ihr}.} This unusual fact is a result of the invariance of the galileon field $\pi$ under the galilean shift symmetry,
\begin{equation}
\pi(x) \to \pi(x)+c+b_\mu x^\mu. \label{eq:galsym}
\end{equation}
Under this transformation, the Lagrangian shifts by a total derivative, leaving the equations of motion unaffected. 

It turns out that single-field galileons do not admit stable solitons \cite{Endlich:2010zj}, the proof of which we will review in \cref{sec:dergal}. One possibility in the hunt for galileon solitons is to extend to theories of multiple galileons, typically possessing some internal symmetry \cite{Padilla:2010de,Padilla:2010tj,Hinterbichler:2010xn,Padilla:2010ir}. For example, a stable solitonic solution akin to a texture has been found in the case of multigalileons with an internal SO(4) symmetry \cite{Padilla:2010ir}. Another is to consider single-field cousins of the galileon, or \emph{generalized galileons}, which maintain second-order equations of motion due to symmetries differing from \cref{eq:galsym}.

Galileons have been generalized in a variety of ways. Due to their wide applicability in cosmology, they have been formulated in curved spacetimes, leading to the \emph{covariant galileons} \cite{Deffayet:2009wt}. Covariantizing the galileons is not simple as it requires the introduction of non-minimal coupling to gravity beyond the cubic term in order to keep the equations of motion second-order. These non-minimal couplings destroy the shift symmetry \eqref{eq:galsym}.

An interesting construction was found in \rcite{deRham:2010eu} which points the way towards a method for systematically constructing generalized galileons. The galileon Lagrangian appears as the non-relativistic limit of the bending mode of a probe brane embedded in a non-dynamical bulk. In this probe-brane construction, only a finite number of actions---specifically the Lovelock invariants and their boundary terms---will lead to second order equations of motion for the bending mode $\pi$. The scalar field $\pi$ inherits its symmetries from a combination of the bulk's Killing symmetries and the brane's reparametrization invariance. For the simplest example, a Minkowski brane embedded in a Minkowski bulk, the Lagrangian for $\pi$ is the \emph{DBI galileon}, and in the small-field limit (corresponding to the non-relativistic limit for the brane) this reduces to the standard galileon. This construction has been generalized to every combination of maximally-symmetric four-dimensional branes and five-dimensional bulks \cite{Goon:2011qf}; to higher co-dimensions \cite{Hinterbichler:2010xn}, leading to multi-galileons; and to cosmological bulks \cite{Goon:2011xf}. Another interesting construction starts by noting that the symmetry~(\ref{eq:galsym}) is nonlinearly realized, and leads to the conclusion that the galileons as Wess-Zumino terms for spontaneously-broken spacetime symmetries \cite{Goon:2012dy}.

This paper is structured as follows. In \cref{sec:gals} we briefly review the galileons and their generalizations. In \cref{sec:dergal} we summarize the no-go theorem for stable, static solitons in galileons and discuss how to evade it. Afterwards, in \cref{sec:zero}, we consider the crucial zero-mode argument from the no-go theorem and check its applicability to generalized galileon theories. In \cref{sec:potential} we consider galileons equipped with a potential. In \cref{sec:move} we consider solitons moving at the speed of light, which avoid many of the arguments made in the preceding sections. We find moving domain-wall-like solutions and analyze their stability. Finally, we conclude in \cref{sec:concl}.

\section{A galileon primer}\label{sec:gals}

Before discussing defect solutions, we will very briefly recap the structure of galileon theories, as well as the ideas behind the generalized theories which we will be investigating. The Lagrangians of the generalized galileons are presented in \cref{app:gglag}.

In four dimensions there are five galileon Lagrangians, labeled $\lag_n$ and given by
\begin{align}\label{eq:gallag}
\lag_1 &= \pi, \nonumber \\
\lag_2 &= (\partial\pi)^2, \nonumber \\
\lag_3 &= (\partial\pi)^2\Box\pi, \nonumber \\
\lag_4 &= (\partial\pi)^2\left[(\Box\pi)^2 - \partial_\mu\partial_\nu\pi\partial^\mu\partial^\nu\pi\right], \nonumber \\
\lag_5 &= (\partial\pi)^2\left[(\Box\pi)^3 - 3\Box\pi\partial_\mu\partial_\nu\pi\partial^\mu\partial^\nu\pi\right. \nonumber \nonumber \\
&\left.\hphantom{{}=(\partial\pi)^2\big[}+ 2\partial_\mu\partial_\nu\pi\partial^\nu\partial_\alpha\pi\partial^\alpha\partial^\mu\pi\right].
\end{align}
Note that $\lag_1$ is a tadpole and $\lag_2$ is a canonical kinetic term (up to a factor of $-1/2$), while the final three are non-trivial higher-derivative terms. These are the unique interaction terms which obey the galilean symmetry, and whose equations of motion are second-order in derivatives. In $N$ dimensions, galileons up to $\lag_{N+1}$ can be constructed.

These properties are easier to see when we note that, after integrations by parts, the galileon Lagrangians can be rewritten (up to constant factors) as
\begin{align}
\lag_2 &\propto \varepsilon^{\mu\cdot\cdot\cdot}\epsilon^{\nu}{}_{\cdot\cdot\cdot}\partial_\mu\pi\partial_\nu\pi, \nonumber \\
\lag_3 &\propto \varepsilon^{\mu\alpha\cdot\cdot}\epsilon^{\nu\beta}{}_{\cdot\cdot}\partial_\mu\pi\partial_\nu\pi\partial_\alpha\partial_\beta\pi, \nonumber \\
\lag_4 &\propto \varepsilon^{\mu\alpha\rho\cdot}\epsilon^{\nu\beta\sigma}{}_{\cdot}\partial_\mu\pi\partial_\nu\pi\partial_\alpha\partial_\beta\pi\partial_\rho\partial_\sigma\pi, \nonumber \\
\lag_5 &\propto \varepsilon^{\mu\alpha\rho\gamma}\epsilon^{\nu\beta\sigma\lambda}\partial_\mu\pi\partial_\nu\pi\partial_\alpha\partial_\beta\pi\partial_\rho\partial_\sigma\pi\partial_\gamma\partial_\lambda\pi,
\end{align}
where dots denote contracted indices. This antisymmetric structure is responsible for many of the notable properties of the galileon. Under the galilean transformation \eqref{eq:galsym}, which implies $\partial_\mu\pi \to \partial_\mu\pi + b_\mu$, with $\partial_\mu b_\nu=0$, these Lagrangians are manifestly shifted by total derivatives due to the presence of the antisymmetric Levi-Civita symbols and the fact that partial derivatives commute. The second-order nature of the equations of motion follows similarly. For example, consider $\lag_3$, although the argument generalizes trivially to the other galileons. After varying the action, the terms which could lead to dangerous higher derivatives in the equations of motion are $\varepsilon^{\mu\alpha\cdot\cdot}\epsilon^{\nu\beta}{}_{\cdot\cdot}b_\mu\partial_\nu\pi\partial_\alpha\partial_\beta\delta\pi$ and $\varepsilon^{\mu\alpha\cdot\cdot}\epsilon^{\nu\beta}{}_{\cdot\cdot}b_\mu b_\nu\partial_\alpha\partial_\beta\delta\pi$. The latter is manifestly a total derivative and will not contribute to the equations of motion. The former, after integration by parts, becomes $\varepsilon^{\mu\alpha\cdot\cdot}\epsilon^{\nu\beta}{}_{\cdot\cdot}b_\mu\partial_\alpha\partial_\beta\partial_\nu\pi\delta\pi$, which vanishes due to symmetry. Furthermore, it is clear that the number of galileons one can construct is limited by the number of indices carried by the Levi-Civita symbol, so that there are $N+1$ galileons in $N$ dimensions.

The galileons were originally discovered in a particular decoupling limit of the higher-dimensional Dvali-Gabadadze-Porrati (DGP) model \cite{Dvali:2000hr}. In this model, the galileon arises as the brane-bending mode of a dynamical four-dimensional brane living in a dynamical five-dimensional bulk. As mentioned earlier, galileons can also be constructed by considering a probe brane embedded in a non-dynamical bulk. Contrary to the DGP model, in the probe brane construction there is no need to take a decoupling limit. The galileon $\pi$ is nothing other than the position of the brane in a specific foliation. The case of a flat brane embedded in a flat bulk leads to the DBI galileons whose precise form is given in \cref{app:gglag}. As an example, the $\lag_2$ term is
\begin{equation}
\lag_2 = \sqrt{1+(\partial\pi)^2},
\end{equation}
which is the standard DBI kinetic term. It is clear that in the small-field limit, $(\partial\pi)^2\ll1$, corresponding to the brane moving and bending at non-relativistic speeds, this reduces to the canonical kinetic term. Indeed, in this limit the $\lag_3$ through $\lag_5$ terms of the DBI galileons reduce to their standard galileon counterparts as well. The DBI galileons are invariant under the symmetry
\begin{equation}
\pi \to \pi + c + b_\mu x^\mu - b_\mu\pi\partial^\mu\pi, \label{eq:DBIsym}
\end{equation}
which in the small-field limit trivially reduces to the galilean symmetry \eqref{eq:galsym}. The physical origin of the galilean symmetry is made clearer through this higher-dimensional origin story as well: \cref{eq:DBIsym} is a combination of bulk Poincar\'e invariance\footnote{Ignoring the unbroken four-dimensional Poincar\'e symmetry.} and brane reparametrization invariance.

This origin of the galileon terms from a probe-brane construction led to a systematic construction of \emph{generalized galileons} for maximally-symmetric probe branes in maximally-symmetric bulks \cite{Goon:2011qf}, so that the brane and bulk can each be Minkowski, de Sitter (dS), or anti-de Sitter (AdS).\footnote{Up to the caveat that dS$_5$ cannot be foliated by AdS$_4$ or M$_4$ slices and M$_5$ cannot be foliated by AdS$_4$ slices; however, these foliations are possible if we consider a bulk with more than one time direction.} These theories, as well as the theories constructed by taking limits (such as the small-field limit or a ``small-derivatives" limit), lead to a rich family of higher-derivative scalar-field theories with second-order equations of motion. The generalized galileons which we study in this paper are presented in \cref{app:gglag}.

In this construction, the conformal galileons arise as the small-derivative limit of the conformal DBI galileons (obtained by considering a Minkowski brane in an AdS bulk). They can also be obtained by promoting the galileon symmetry to a conformal one; that is, promoting the Poincar\'e and Galilean group to the conformal SO($4,2$) group. The conformal galileon is invariant under dilations and a special conformal transformation that read
\begin{align}
\delta\pi&=1-x^\mu\partial_\mu\pi , \\
\delta_\mu\pi&=2x_\mu+x^2\partial_\mu\pi-2x_\mu x^\nu\partial_\nu\pi.
\end{align}
On the other hand, the (A)dS galileons are found by taking the small field limit of the theories where a maximally symmetric brane is embedded in a (A)dS bulk.

In addition to these galileons constructed from maximally-symmetric probe branes and bulks, we will also consider an interesting and very natural generalization of the galileons known as \emph{covariant galileons} \cite{Deffayet:2009wt}. These are motivated by the fact that the standard galileons live on a flat background, so for curved spacetimes we need some covariantization procedure. A natural approach to covariantizing the galileons is to simply couple them minimally to the metric by promoting $\partial\to\nabla$. While this is adequate for $\lag_2$ and $\lag_3$, this generates higher derivatives of the metric for $\lag_4$ and $\lag_5$. This issue can be remedied by adding specific non-minimal curvature couplings in these terms which keep the equations of motion for both $\pi$ and the metric $g_\mn$ second-order. These non-minimal couplings are presented in \cref{eq:covgals}.

\section{Derrick's theorem and the zero-mode argument} \label{sec:dergal}

When dealing with canonical kinetic terms in a scalar field theory in $d>1$ dimensions, Derrick's theorem \cite{Derrick:1964ww} tells us that there are no stable, stationary solitons. This is shown by assuming the existence of a stable soliton $\phi_0(x)$; perturbing this solution by a spatial dilation $x \rightarrow \lambda x$, it is straightforward to show that there are no stable stationary points of the energy with respect to $\lambda$. It is well-known that Derrick's theorem can be evaded by including a gauge field or considering a time-dependent solution. A variation of this theorem was proved in \rcite{Endlich:2010zj} for the case of a single galileon. The argument therein requires the existence of a zero mode of the perturbations and demonstrates the non-existence of stable solitonic solutions in galileons in $d$ dimensions. Here we will briefly sketch this proof.

Consider a theory of a single galileon field with a Lagrangian $\lag=\sum_nc_n\lag_n$, where the galileon Lagrangians $\lag_n$ are given in \cref{eq:gallag}. Let us assume there exists a soliton solution $\pi_0(\mathbf{x})$ and consider small perturbations $\varphi$ about it. The energy of the fluctuations is given by
\begin{equation}
\delta E=\frac{1}{2}\int \ud^dx Z^{ij}(\mathbf{x})\partial_i\varphi\partial_j\varphi, \label{de}
\end{equation} 
where
\begin{equation}
Z^{ij}(\mathbf{x})=c_2\delta^{ij}+c_3\left(\delta^{ij}\nabla^2\pi_0-\partial^i\partial^j\pi_0\right)+\cdots. \label{z}
\end{equation}
In order to have a stable solution, we require $Z^{ij}(\mathbf{x})$ to be positive semi-definite everywhere in space. From \cref{z} we see that, far from the soliton's core, the $c_2$ term dominates, implying that $c_2>0$.

The key point in this proof is the existence of zero modes; perturbations with $\delta E=0$. Because the galileons are invariant under spatial translations, the energy of the soliton is clearly unaffected by translations, $\varphi_\epsilon=\epsilon\cdot\nabla\pi_0$. The existence of this zero mode implies that $Z^{ij}(\mathbf{x})$ must have a negative eigenvalue in some region of space in order to compensate for the positive $c_2$ term. This eigenvalue signals a gradient instability for the fluctuations, so that if the soliton exists it must be unstable. These kind of instabilities can affect short-wavelength modes leading to fast decay rates, and so we would not be able to trust the effective field theory.

There are several ways to circumvent the zero-mode argument. One possibility is the existence of a potential for the perturbations; however, this will break the generalized galileon symmetry in some cases. Another is to introduce time dependence, and a further route is to consider multiple fields. In the first case, the existence of an extra potential term relaxes the requirement of having a negative eigenvalue of $Z^{ij}$ in some region in space in order to have a vanishing energy for the zero mode. By introducing time dependence, we remove one of the main assumptions in the previous proof, as we required the field to be static. In particular, this argument implies that one should investigate solitons that move at the speed of light, as there is no frame in which such a solution is static. This possibility was investigated in \rcite{Masoumi:2012np}, where it was shown that for a single-field galileon, a localized lump traveling at the speed of light in one direction is stable against small perturbations. In the case of multi-galileons, the structure of the energy for the perturbations is more complicated and the crucial zero-mode argument no longer holds. For the case of a SO(4) multi-galileon confined on a $S^3$, a soliton solution exists, and while its stability has not been proven, it is expected from topological arguments \cite{Padilla:2010ir}. Having a theory confined to a sphere is equivalent to adding a constraint $\lambda (|\pi|^2-1)$ to the Lagrangian, which leads to a potential term in the Lagrangian for $\varphi$, rendering the zero-mode argument invalid. 

\section{Zero-mode argument in generalized galileons} \label{sec:zero}
\subsection{DBI galileons}

In this section, we investigate the extent to which the no-go theorem of \rcite{Endlich:2010zj} described in the previous section---relying crucially on the zero-mode argument---can yield information about the existence (or lack thereof) of stable solitons in the generalized galileon theories discussed in \cref{sec:gals}. In some cases this is a rather obvious extension of the original results, whereas in others, interesting obstacles exist. 

We start with the DBI galileons \cite{deRham:2010eu} whose Lagrangians are given in \cref{eq:DBIgals}. As before, we assume that there exists a soliton solution $\pi_0(\mathbf{x})$. In order to calculate the energy of the fluctuations we use the fact that given the action
\begin{equation}
S_\varphi=\frac{1}{2}\int \ud^dx Z^{\mu\nu}(\mathbf{x})\partial_\mu\varphi\partial_\nu\varphi,
\end{equation}
if we perturb the equations of motion of the soliton solution we get the equations of motion for $\varphi$
\begin{equation}
\frac{\delta S_\varphi}{\delta \varphi}=\partial_\mu\left(Z^{\mu\nu}(\mathbf{x})\partial_\nu\varphi\right)=0,
\end{equation}
and from these we can read off the matrix $Z^{\mu\nu}(\mathbf{x})$. We can then analyze the stability by looking at the spatial components,
\begin{align}
Z^{ij}(\mathbf{x})&=\gamma^3\Big[c_2\left(1+(\partial\pi_0)^2\right)\delta^{ij}-c_2\partial^i\pi_0\partial^j\pi_0+\gamma c_3\Big(2\Box\pi_0\left(1+(\partial\pi_0)^2\right)\delta^{ij}\\
&\hphantom{{}=\gamma^3\Big[}-2\left(1+(\partial\pi_0)^2\right)\partial^i\partial^j\pi_0-2\partial^j\pi_0\left(\Box\pi_0\partial^i\pi_0-\partial_k\partial^i\pi_0\partial^k\pi_0\right)\Big)+\cdots\Big].
\end{align}
In a similar way to the galileon case, for regions far from where the soliton is localized the term $c_2\delta^{ij}$ dominates, and so $Z^{ij}$ is positive far from the soliton. Additionally, this theory has a zero mode given by the spatial translation of the soliton. Thus we arrive at the same conclusion as for the standard galileons: if a soliton solution exists, it must be unstable.

\subsection{Covariant galileons}
Now we consider the covariant galileons \cite{Deffayet:2009wt} in a non-dynamical spacetime, with Lagrangians given in \cref{eq:covgals}. Note that the $\lag_2$ and $\lag_3$ are the same as for the flat space galileon with $\partial\rightarrow\nabla$, while the other terms require additional non-minimal couplings to $g_\mn$ in order to maintain second-order equations of motion. These new terms break the galilean symmetry. We assume a solitonic solution exists and analyze its stability by looking at $Z^{ij}(\mathbf{x})$, which in this case is given by
\begin{align}
Z^{ij}(\mathbf{x})&=c_2\delta^{ij}+c_3\left(\nabla^{ij}\nabla^2\pi_0-\nabla^i\nabla^j\pi_0\right) +\cdots \\
&\hphantom{{}=}+c_4 R_0\left(\nabla_k\pi_0\nabla^k\pi_0\delta^{ij}+2\nabla^i\pi_0\nabla^j\pi_0\right)+\cdots.
\end{align}
Since the soliton is a localized solution, we could expect the spacetime to be flat far from the core of the soliton. If this is the case then, far from where the soliton is localized, the $c_2$ term dominates; even if the space is asymptotically (A)dS, this term will still dominate as long as the gradients vanish fast enough relative to the curvature terms. 

Next, we should worry about the existence of zero modes. Previously we used translation modes since Minkowski space is translationally invariant. For the covariant galileon, translations will provide a zero mode if the spacetime is spatially homogeneous. In this case, we can apply the zero-mode argument as in galileons and DBI galileons. We find that stable, static solitons in covariant galileons living in a spatially homogeneous space are ruled out.

\subsection{Conformal and (A)dS galileons: no zero-mode argument}

Interestingly, there are a few examples of generalized galileons, constructed using probe branes, for which the zero-mode argument fails due to the presence of a potential for the perturbations in the energy functional.\footnote{Note that this does not prove the existence of solitons, but rather signals the absence of a powerful argument which might have been used to prove their \emph{nonexistence}.} This is the case in particular for the conformal, dS, and AdS galileons, in both their DBI versions (which come directly from the probe-brane construction) and the small-field or small-derivative limits that yield analogues of the galileons.\footnote{In the language of \rcite{Goon:2011qf}, the results in this section apply to the AdS DBI, conformal DBI, and Type III dS DBI galileons---all of which are the same theory in different slicings, although their limiting theories differ---as well as to the Type I dS DBI galileons.} The corresponding Lagrangians for all of these theories can be found in \rcite{Goon:2011qf}, while we present the Lagrangians for the limiting theories in \cref{app:gglag}.

As before, we consider a solitonic solution $\pi$ and its perturbations $\varphi$. The energy functional in each of these theories can be written in the form
\begin{equation}
\delta E=\frac{1}{2}\int \ud^dx \left[ Z^{\mu\nu}(\mathbf{x})\partial_\mu\varphi\partial_\nu\varphi+V(\pi,(\partial\pi)^n)\varphi^2\right].
\end{equation}
For the conformal DBI galileons, each term contributes to the potential. For example, $\lag_1=-\frac{R}{4}e^{-4\pi/R}$ will clearly give such a contribution, while from $\lag_2$ we get a term $V\supset\frac{8}{\mathcal{R}} e^{-\frac{4 \pi_0}{\mathcal{R}}} \gamma^{-1}$, where $\gamma\equiv1/\sqrt{1+ e^{\frac{2 \pi_0}{\mathcal{R}}} (\partial\pi_0)^2}$. Meanwhile, the (A)dS DBI galileons have explicit potential terms that will contribute to $V(\pi,(\partial\pi)^n)$.

Let us consider the behavior of $\varphi$ far from the core of the soliton. As before, we demand that the eigenvalues of $Z^{ij}$ be positive and non-zero. If we use the zero-mode argument, we do not find that a negative eigenvalue of $Z^{ij}$ in another region of space is required for the energy of the perturbations to vanish, as there are extra potential terms. Thus, it is clear that we are not be able to reach any conclusion on the stability or existence of solitons in these generalized galileons.

\section{Generalized galileons with a potential} \label{sec:potential}
We have seen that the existence of a potential term helps us circumvent the zero-mode argument, opening up the possibility of stable solitonic solutions with generalized galileons. Another important check is whether a scaling argument rules out stable solutions, as it does for the standard galileons\footnote{The standard spatial dilation argument does not rule out the existence of stable solitons thanks to the non-canonical kinetic terms.}. Here we will show that, as might be expected, having a symmetry-breaking potential is enough to evade this argument. We will analyze various scenarios for generalized galileons with a potential that could give rise to domain walls.

\subsection*{A note on the scaling argument} \label{note}
If we assume that a soliton solution exits for galileons and we rescale the amplitude of the solution as $\pi(x)\rightarrow\pi_{\omega}(x)=\omega\pi(x),$
the energy functional will be given by
\begin{equation}
E(\omega)=\sum_n E_n(\omega) = \sum_n \omega^n E_n^{(0)},
\end{equation}
where the $E_n$ correspond to the terms with $n$ copies of the field and the superscript $(0)$ denotes the energy when $\omega=1$. Note first a simple result: if all the $E_n^{(0)}$ are positive, then clearly we can always reach lower-energy solitons by choosing smaller values of $\omega$. Since in this case we could always perturb the solution to decrease its energy further, such a soliton would not be stable.

Let us analyze the case of a scalar field with a polynomial Lagrangian with terms up to quartic order. The energy functional (dropping the superscript $(0)$) is given by
\begin{equation}
E(\omega)=\omega E_1+\omega^2 E_2 +\omega^3 E_3 +\omega^4 E_4.
\end{equation}
Demanding that the soliton extremize the energy and be stable implies that
\begin{align}
\left.\frac{\ud E}{\ud \omega}\right|_{\omega=1}&=E_1+2E_2+3E_3+4E_4=0, \label{e1}\\
\left.\frac{\ud^2 E}{\ud \omega^2}\right|_{\omega=1}&=-(3E_1+4E_2+3E_3)>0,
\end{align}
where in the second line we have solved for $E_4$ using \cref{e1}. Splitting $E_2$ into its kinetic and potential contributions, $E_2 \equiv K_2 + V_2$, and assuming that the kinetic part, which corresponds to the canonical kinetic term, is positive, we have that
\begin{equation}
0<K_2<-\frac{3}{4}(E_1+E_3)-V_2. \label{stw}
\end{equation}
This condition should be satisfied by a stable solution, but does not prove the existence of such a solution.

Let us consider the specific case of a symmetry-breaking potential which has a negative quadratic term. In \cref{stw}, we have a negative contribution from the kinetic terms, but the potential term will give an infinite\footnote{ This infinite contribution is not a problem since it is canceled by the other potential terms when considering the total energy.} positive contribution, which causes the stability condition to always be satisfied. This means that the scaling argument does not rule out stable, static solitons in a scalar field theory with a polynomial Lagrangian and a symmetry-breaking potential. This motivates us to look for solitons in galileons with a symmetry-breaking potential.

\subsection{Galileons}
If we add a potential to the standard galileon Lagrangian then soliton solutions can be found. Of course, the potential itself breaks the galilean symmetry, but nevertheless, this example provides a simple playground in which to explore the effects of galileon-type terms on soliton solutions. For simplicity, let us focus on the cubic galileon,
\begin{equation}
\lag=-\frac{1}{2}(\partial\pi)^2(1+\frac{1}{\Lambda^3}\Box\pi)-V(\pi),
\end{equation} 
with a symmetry-breaking potential
\begin{equation}
V(\pi)=\lambda(\pi^2-v^2)^2.\label{eq:symbreakpot}
\end{equation}

For a system which only depends on one spatial dimension---which we will take, without loss of generality, to be the Cartesian coordinate $z$---it is always possible to calculate the first integral of the equations of motion due to the fact that momentum is conserved in this direction. A simple way of obtaining this first integral is to calculate the conserved charge
\begin{equation}
J=\pi'\frac{\partial\lag}{\partial\pi'}-\lag+\frac{\partial\lag}{\partial\pi''}\pi''-\frac{\ud}{\ud z}\frac{\partial\lag}{\partial\pi''} \pi',
\end{equation}
where $'\equiv\ud/\ud z$. The fact that $J$ is conserved follows from the Euler-Lagrange equations. For the cubic galileon, $J$ turns out to be the same as for a canonical scalar field,
\begin{equation}
J=-\frac{1}{2}\pi'^2+V(\pi),
\end{equation} 
which means that we have the same domain walls as in the canonical case. This is expected since we effectively have a one-dimensional problem and thus the galileon terms are total derivatives. 

\subsection{DBI galileons} 
Now let us consider equipping the cubic DBI galileons with a potential in order to look for domain walls. The Lagrangian is given by a combination of $\lag_2$ and $\lag_3$ in \cref{eq:DBIgals},
\begin{equation}
\lag=-\sqrt{1+(\partial\pi)^2}-a\left(\Box\pi+\frac{\partial_\mu\pi\partial_\nu\pi\partial^\mu\partial^\nu\pi}{\sqrt{1+(\partial\pi)^2}}\right)-V(\pi)
\end{equation} 
along with the symmetry-breaking potential \eqref{eq:symbreakpot}. Taking the field to depend only on the coordinate $z$ we find that the conserved charge $J$ is the same as for a DBI scalar,
\begin{equation}
J=-\frac{1}{\sqrt{1+\frac{1}{2}\pi'^2}}+V(\pi).
\end{equation} 
This fact again follows from the new terms being total derivatives. This implies that we obtain the same results as the DBI case~\cite{Babichev:2006cy}. Such walls can have interesting effects not present in the canonical case. For example, when constructing a domain wall in DBI, there exist values of the parameters for which a singularity develops, and forbids the existence of a global solution. In the language of effective field theories, near this singularity the theory is strongly-coupled which obscures the validity of these results \cite{Andrews:2010eh}.

\subsection{Conformal galileons}

As in the previous cases, we include a symmetry-breaking potential to the conformal galileons with Lagrangian is given by \cref{eq:confgals}, in order to find domain wall solutions. We will consider the Lagrangian up to the cubic term and assume that the field only depends on the Cartesian coordinate $z$, so that the Lagrangian reads
\begin{equation}
\lag=-\frac{1}{2}\pi'^2 e^{-2\pi }\left[1+a\,e^{2\pi }\left(\pi''-\frac{1}{4}\pi'^2\right)\right]-\lambda(\pi^2-v^2)^2.
\end{equation}
The corresponding conserved charge is given by
\begin{equation}
J=-\frac{1}{2}\pi'^2 e^{-2\pi }\left[1+\frac{3}{2}a\,e^{2\pi }\pi'^2\right]+\lambda(\pi^2-v^2)^2=0. \label{jconf}
\end{equation}
It is straightforward to see that solving this equation with the boundary condition $\pi'(\infty)=0$ and $\pi(\infty)=v$ results in an imaginary derivative. From \cref{jconf} we can solve algebraically for $\pi'$, picking the branch that gives the desired boundary conditions gives
\begin{equation}
\pi'^2=\frac{\sqrt{12 a \lambda  \left(\phi ^2-v^2\right)+e^{-4  \phi }}-e^{-2  \phi
	}}{3\, a }.
\end{equation}
Requiring a positive value inside the square root gives the constraint $a<\frac{e^{-4  v}}{12 \lambda  v^2}$. Even if this constraint is satisfied, $\pi'^2$ will always be negative. This indicates that the solution does not exist. We conclude that there are no domain wall solutions for cubic conformal galileons with a symmetry-breaking potential.
	
\subsection{(A)dS galileons}

The dS and AdS galileons, as their names indicate, live in dS and AdS space, respectively. They have the interesting feature of naturally possessing polynomial potentials, as can be seen in \cref{eq:AdSgals}. This is in sharp contrast to the previous examples, where by adding in a potential we broke the (generalized) galilean symmetry. If we impose a $\mathbb{Z}_2$ symmetry only the $\lag_2$ and $\lag_4$ terms remain, which have $\pi^2$ and $\pi^4$ potentials, respectively. This yields a symmetry-breaking potential of the form \eqref{eq:symbreakpot} (as long as there is a relative sign between $\lag_2$ and $\lag_4$), which suggests the presence of domain walls. Given this, we take the Lagrangians to be
	\begin{align}
	\text{dS:}&\quad\quad\lag_2-a\lag_4\\
	\text{AdS:}&\quad-\lag_2+a\lag_4.
	\end{align}
We may write the potential for both cases in a unified form,
	\begin{equation}
	V(\pi)=\frac{|R|}{288}\left(-48\pi^2+aR^2\pi^4\right)+\frac{2}{a|R|},
	\end{equation}
where we have added an extra constant term so that the potential vanishes at the minimum. Note that, with this choice, we will have a ghost around $\pi=0$ for the AdS case and around the minimum $\pi=\pm\sqrt{24/a}/R$ for the dS case \cite{Goon:2011qf}.

In order to construct a domain-wall solution we need to pick the coordinates of the spacetime and the orientation of the wall. One possibility is to work in global coordinates for (A)dS, where the metric is given by
	\begin{equation}
	\ud s^2=-f(r) \ud t^2+f(r)^{-1} \ud r^2+r^2\ud \Omega_2^2,\qquad f(r)=1-\frac{r^2}{12}R.
	\end{equation}
	In this case, the field configuration is a bubble: $\pi$ depends only on the coordinate $r$, and the appropriate boundary conditions are
	\begin{equation}
	\pi(\infty)=\frac{1}{R}\sqrt{\frac{24}{a}},\quad\pi(0)=-\frac{1}{R}\sqrt{\frac{24}{a}}. \label{bcb}
	\end{equation}
	
This construction faces several problems. The energy inside and outside the bubble is the same, so we expect the surface tension of the wall to cause it shrink. This is the case for canonical kinetic terms, which provide a positive surface tension. One might hope that galileon terms give a different result; the contribution from the second derivative terms becomes negative on one side of the wall and might dominate the energy density causing the total surface tension to vanish. Besides this issue, we face the problem of the field becoming ghostly. The boundary conditions for the bubble are those in \cref{bcb}, so $\pi(R)=0$​ for some $r=R$​. We also know that around $\pi=0$ for the AdS galileons and $\pi=\pm\sqrt{24/a}/R$ for dS galileons the field becomes a ghost. This means that in both cases the soliton solution will become ghostly at some $r$.	

\section{Moving solitons} \label{sec:move}
Another possible way of avoiding Derrick's theorem is to include time dependence. For the standard galileon model, it has been shown~\cite{Masoumi:2012np} that solitons traveling at the speed of light can evade the zero-mode argument, and that these domain-wall like solutions are stable under small perturbations. In this section we will investigate the stability of solitons moving at the speed of light for DBI and conformal galileons. We will not consider the (A)dS galileons here, since the presence of potential terms implies that solitons moving at the speed of light are not solutions of their equations of motion.

\subsection{DBI galileons}

We start with the DBI galileons \eqref{eq:DBIgals} and consider only the $\lag_2$ and $\lag_3$ terms for simplicity. The Lagrangian for a general combination of these two terms is
\begin{equation}
\lag = -\sqrt{1+(\partial\pi)^2} +a\left(-\Box\pi+\gamma^2\partial_\mu\pi\partial_\nu\pi\partial^\mu\partial^\nu\pi\right),
\end{equation}
where $\gamma=1/\sqrt{1+(\partial\pi)^2}$, and the equations of motion are
\begin{align}
\Box\pi\left(\frac{1}{\sqrt{1+(\partial\pi)^2}}+a\frac{\partial_\nu\partial^\nu\pi}{1+(\partial\pi)^2}\right)-\frac{\partial^\mu\pi\partial_\nu\partial_\mu\pi\partial^\nu\pi}{(1+(\partial\pi)^2)^{3/2}}&\nonumber\\
+a\left(\frac{2 \partial^\mu\pi\partial^\nu\pi\partial_\lambda\partial_\nu\pi\partial^\lambda\partial_\mu\pi}{(1+(\partial\pi)^2)^{2}}-\frac{2 \partial^\mu\pi\partial_\nu\partial_\mu\pi\partial^\nu\pi \partial_\lambda\partial^\lambda\pi }{(1+(\partial\pi)^2)^{2}}\right)&=0.
\end{align}
We will work in lightcone coordinates, $u=x^0+x^1$ and $v=x^0-x^1$, and assume $\pi=\pi(u,v)$ so that the system reduces to a $(1+1)$-dimensional one. In lightcone coordinates the equations of motion read
\begin{align}
2 a \sqrt{1+2 \partial_v\pi \partial_u\pi}\;
\partial_u\partial_v\pi^2-\partial^2_v\pi \left(2 a
\sqrt{1+2 \partial_v\pi \partial_u\pi}\; \partial^2_u\pi+2 \partial_v\pi \partial_u\pi^3+\partial_u\pi^2\right)& \nonumber\\
-\left(2 \partial_v\pi \partial_u\pi+1\right) \partial^2_u\pi \partial_v\pi^2+\left(4 \partial_v\pi^2 \partial_u\pi^2+6 \partial_v\pi \partial_u\pi+2\right) \partial_u\partial_v\pi&=0.
\end{align}
It is clear that if $\pi$ is a function of $u$ or of $v$ alone, then the equations of motion are satisfied. In fact, this will be the case for any single field in flat space with no potential term: whenever we contract indices in lightcone coordinates, the structure of the metric---$g_{uu}=g_{vv}=0$---forbids terms involving only $u$ or only $v$ components. This allows us to set initial conditions with a localized lump and let it propagate to the right or left in the $x^1$ direction at the speed of light; this is the \emph{moving soliton}.

\subsubsection{Stability}
We analyze the stability of this moving soliton by perturbing the DBI galileon equations of motion to linear order. Defining
\begin{equation}
\pi(u,v) = \pi_\rb(u) + \phi(u,v),
\end{equation}
where $\pi_\rb$ is the moving soliton found above, which we have taken without loss of generality to be a left mover, $\pi_\rb=\pi_\rb(u)$, the linearized equation of motion is
\begin{equation}
2 \partial_u\partial_v\phi-f(\pi_\rb)\partial^2_v\phi=0, \label{phieom}
\end{equation}
where $f(\pi_\rb)=2 a \pi _\rb''+\pi _\rb'^2$. To simplify the stability analysis, we expand the perturbation in a complete set of functions
\begin{equation}
\phi=\sum_n A_n e^{-i(k_n x-\omega_n t)} \ ,
\end{equation}
which yields for a given mode (removing the subscript $n$)
\begin{equation}
k^2 \left(-2+f(\pi_\rb)\right)+2k\omega f(\pi_\rb)+ \omega ^2 \left(2+f(\pi_\rb)\right)=0.
\end{equation}
Solving for $\omega$ we find
\begin{equation}
\omega=-k \qquad \omega=-k\frac{f(\pi_\rb)-2}{f(\pi_\rb)+2}.
\end{equation}
We see that if the background solution is real-valued, then $\omega$ is real for any (real) $k$. These solutions oscillate rather than growing exponentially, and are therefore stable.

Alternatively, we can analyze the \textit{local} stability, as defined in \rcite{Nicolis:2004qq}, at the level of the action. In order to do so, we will assume that the characteristic time and length scales of the perturbations are much smaller than the typical scales on which the background solution varies, so that we may treat $\pi_\rb$ and its derivatives as constants. The quadratic Lagrangian for $\phi$ is
\begin{equation}
\lag_\phi=\frac{1}{2}Z^{\mu\nu}\partial_\mu\phi\partial_\nu\phi,
\end{equation}
where, per \cref{phieom}, $Z^{\mu\nu}$ is given by
\begin{equation}
Z^{\mu\nu}=-g^{\mu\nu}+f(\pi_\rb)\delta^\mu_v\delta^\nu_v.
\end{equation}
Diagonalizing this we find,
\begin{equation}
Z^{\mu\nu}=
\begin{pmatrix}
\frac{1}{2} \left(f(\pi_\rb)-\sqrt{\left(f(\pi_\rb)\right){}^2+4}\right) & 0 &0 &0 \\
0 & \frac{1}{2}
\left(f(\pi_\rb)+\sqrt{\left(f(\pi_\rb)\right){}^2+4}\right)& 0 &0\\
0 &0 & 1&0\\
0&0&0&1
\end{pmatrix}.
\end{equation}
We can see that one of the eigenvalues is positive and the other one is negative, so that $Z^\mn$ has the healthy signature $(-,+,+,+)$. 

Another way of seeing the local stability condition is by requiring all the eigenvalues of $Z^\mu_\nu$ to be negative. The matrix $Z^\mu_\nu$ reads 
\begin{equation}
Z^{\mu}_\nu=
\begin{pmatrix}
-1 & f(\pi_\rb)& 0 &0 \\
0 & -1& 0 &0\\
0 &0 & -1&0\\
0&0&0&-1
\end{pmatrix},
\end{equation}
so that the eigenvalues will indeed be negative. An issue to notice here is that although $Z^\mu_\nu$ cannot be diagonalized, the analysis at the level of the equations of motion indicates stability.  We conclude that a lump of DBI galileon moving at the speed of light is locally stable against small fluctuations.

The analysis above does not account for ghost-like instabilities related to the global sign of the Lagrangian. If $\pi$ interacts with other matter, then we also need to ensure that $Z^{00}>0$ in order to avoid ghost instabilities. This translates to 
\begin{equation}
Z^{00}=\frac{1}{2}+\frac{1}{4}f(\pi_\rb)>0.
\end{equation}

\subsubsection{Energy} It is possible to covariantize the DBI galileons up to $\lag_3$ by replacing partial derivatives with covariant ones while keeping the equations of motion second-order. By doing this we can obtain the stress-energy tensor,
\begin{align}
T_{\mu\nu} &= \gamma\nabla_\mu\pi\nabla_\nu\pi -\gamma^{-1}g_{\mu\nu} +a\gamma^2\left(\Box\pi\nabla_\mu\pi\nabla_\nu\pi -2 \nabla_{(\mu}\pi\nabla_{\nu)}\nabla_\alpha\pi\nabla^\alpha\pi + \nabla_\alpha\pi\nabla_\beta\pi\nabla^\alpha\nabla^\beta\pi g_{\mu\nu}\right),
\end{align}
where we have defined $\gamma\equiv(1+(\partial\pi)^2)^{-1/2}$. Without loss of generality, we assume a left moving solution $\pi_\rb=\pi_\rb(u)$ and find that the energy is given by
\begin{equation}
E=\int \ud^3xT^{00}=\int\ud^3x\left(1+\pi'^2_\rb\right). \label{mdbie}
\end{equation}
This energy is positive and does not depend on the higher order DBI galileon terms: it is the same energy as in the DBI case $a=0$. This feature was also found for the standard galileons. From \cref{mdbie} we can then see that steeper lumps have larger energy.

\subsection{Conformal galileons}

Finally, we consider the conformal galileons \eqref{eq:confgals}. We will again take a combination of $\lag_2$ and $\lag_3$,
\begin{equation}
\lag = -\frac{1}{2}  e^{-2 \pi  } (\partial\pi)^2+a\left(-\frac{1}{2} (\partial\pi)^2\Box\pi +\frac{1}{4}(\partial\pi)^4\right),
\end{equation}
for which the corresponding equation of motion is
\begin{align}
e^{-2 \pi }\left(\Box\pi- (\partial\pi)^2\right)+a\left[(\Box\pi)^2-\partial_\mu\partial_\nu\pi\partial^\mu\partial^\nu\pi-\Box\pi(\partial\pi)^2 -2\partial_\mu\pi\partial_\nu\pi\partial^\mu\partial^\nu\pi\right]=0.
\end{align}
Working, as before, in lightcone coordinates, and assuming $\pi=\pi(u,v)$, the equations of motion read
\begin{align}
 e^{-2\pi} \left(\partial_u\partial_v \pi- \partial_u \pi \partial_v \pi\right) - a\left[ \partial^2_v \pi \partial_u \pi^2 +4  \partial_v \pi \partial_u \pi \partial_u\partial_v \pi - (\partial_u\partial_v \pi)^2 + \partial_v \pi^2 \partial^2_u\pi + \partial^2_v \pi \partial^2_u\pi\right]=0.
\end{align}
This admits a solution that only depends on $u$ or $v$, so, as in the DBI galileon case, we can construct a solitonic lump moving at the speed of light.

\subsubsection{Stability}
Perturbing around a background solution $\pi(u,v)=\pi_\rb(u)+\phi(u,v)$ to linear order in the perturbations gives
\begin{equation}
e^{-2 \pi_\rb} \partial_u\partial_v\phi- f(\pi_\rb) \partial^2_v\phi- e^{-2 \pi_\rb} \pi_\rb' \partial_v\phi=0, \label{confeom}
\end{equation}
where $f(\pi_\rb)=a( \pi _\rb''+\pi _\rb'^2)$. Note that in addition to terms like those in the DBI case \eqref{phieom}, we have an extra term that is linear in $\partial\phi$. We  expand the perturbation in terms of plane waves, yielding, for a given mode,
\begin{align}
\omega^2 \left(e^{-2 \pi_\rb}+f(\pi_\rb) \right)+2 i e^{-2 \pi_\rb} \omega  \pi_\rb'+k^2 \left(-2 e^{-2 \pi_\rb}+ f(\pi_\rb) \right) +k \left(2 i e^{-2 \pi_\rb} \pi_\rb' +2 \omega  f(\pi_\rb)\right)=0,
\end{align}
and solving for $\omega$ we find
\begin{equation}
\omega=-k, \qquad \omega=\frac{k-f(\pi_\rb) e^{2 \pi_\rb} k -2i \pi_\rb'}{1+f(\pi_\rb) e^{2 \pi_\rb}}.
\end{equation}
The latter mode has an imaginary part which could lead to exponential growth. This mode behaves like $\propto e^{2\pi_\rb't/(1+f(\pi_\rb) e^{2 \pi_\rb})} g(\pi_\rb)$, where $g$ is an oscillatory function. This mode grows exponentially if 
\begin{equation}
\frac{2\pi_\rb'}{1+f(\pi_\rb) e^{2 \pi_\rb}}>0,
\end{equation}
and otherwise decays exponentially.

The term that appears in the equation of motion with a single derivative is responsible for the imaginary part in the dispersion relation. It comes from the first term in the Lagrangian, which contributes the following term to the equation of motion
\begin{equation}
\partial_u\left(e^{-2 \pi_\rb}\partial_v\phi\right).
\end{equation}

Note that we would not have found this instability if we had treated the background field as a constant as in the local stability analysis. This can be seen by performing the local analysis at the level of the action. In this case $Z^{\mu\nu}$ is given by
\begin{equation}
Z^{\mu\nu}=-e^{-2\pi_\rb}g^{\mu\nu}+f(\pi_\rb)\delta^\mu_v\delta^\nu_v.
\end{equation}
Diagonalizing $Z^{\mu\nu}$ we find (again writing only the non trivial components)
\begin{equation}
Z^{\mu\nu}=
\begin{pmatrix}
\frac{1}{2} \left(f(\pi_\rb)-\sqrt{\left(f(\pi_\rb)\right){}^2+4e^{-4\pi_\rb}}\right) & 0 & 0 &0\\
0 & \frac{1}{2}
\left(f(\pi_\rb)+\sqrt{\left(f(\pi_\rb)\right){}^2+4e^{-4\pi_\rb}}\right)& 0 &0\\
0 &0 & 1&0\\
0&0&0&1
\end{pmatrix}.
\end{equation}
In order to have stability, we need this matrix to have the healthy signature $(-,+,+,+)$ which is clearly satisfied. Another way of seeing local stability is by looking at the matrix $Z^\mu_\nu$ which in this case reads
\begin{equation}
Z^{\mu}_\nu=
\begin{pmatrix}
-e^{-2\pi_\rb} & f(\pi_\rb) & 0 &0\\
0 & -e^{-2\pi_\rb}& 0 &0\\
0 &0 & -e^{-2\pi_\rb}&0\\
0&0&0&-e^{-2\pi_\rb}
\end{pmatrix}.
\end{equation}
We see that the eigenvalues will be negative, as is required for a locally stable solution. As before, it is important to notice that $Z^\mu_\nu$ cannot be diagonalized. However, again, we have shown at the level of the equations of motion that the solution will be unstable if $\frac{2\pi_\rb'}{1+f(\pi_\rb) e^{2 \pi_\rb}}>0$.

\subsubsection{Energy} Similar to the standard galileons and DBI galileons we can covariantize the conformal galileons up to $\lag_3$ by replacing partial derivatives with covariant ones while keeping the equations of motion second-order. The stress-energy tensor is given by
\begin{align}
T_{\mu\nu}&= e^{-2 \pi } \left(\nabla_\mu\pi\nabla_\nu\pi - \frac{1}{2} (\partial\pi)^2g_{\mu\nu}\right) \nonumber \\
&\hphantom{{}=}+a\left[\left(\Box\pi -(\partial\pi)^2\right) \nabla_\mu\pi \nabla_\nu\pi -2 \nabla_{(\mu}\pi\nabla_{\nu)\alpha}\pi\nabla^\alpha\pi +\left(\frac{1}{4} (\partial\pi)^4 + \nabla_\mu\nabla_\nu\pi \nabla^\mu\pi \nabla^\nu\pi\right)g_{\mu\nu}\right]
\end{align}
Assuming a left mover $\pi_\rb=\pi_\rb(u)$, the energy is given by
\begin{equation}
E=\int \ud^3xT^{00}=\int\ud^3x\left(e^{-2\pi_\rb}\pi_\rb'^2\right).
\end{equation}
As with the galileon and DBI galileon cases, this is the same energy as there would be in the absence of non-trivial derivative interactions (i.e., with $a=0$). The energy is positive and indicates that a steeper lump has a larger energy. An interesting feature to notice here is that, due to the exponential factor, a tall lump will have a small energy.

\section{Conclusions}\label{sec:concl}

Generalized galileon theories possess a rich non-linear structure that could sustain solitonic solutions. Despite the no-go theorem that rules out stable, static solitons for the simplest galileons, it is interesting to explore how to circumvent this proof and to explore the types of generalized galileons to which it applies. In this paper, we have explored two ways of avoiding this no-go theorem; one consists of having a potential in the  Lagrangian and the other one considers time-dependent solitons moving at the speed of light. 

We have found that the no-go theorem, which relies on the existence of a zero mode, can be applied to DBI galileons as well as covariant galileons. In the latter case, we were able to rule out the existence of stable, static solitons in covariant galileons living in a spatially homogeneous background. For the case of conformal and (A)dS galileons in both their DBI versions and their small-field or small-derivative limit, the zero mode argument does not apply due to the presence of a potential for the perturbations.

We have exhaustively analyzed the possibility of finding solitons in generalized galileon theories with a potential. Scaling arguments that could rule out stable solutions are easily evaded by theories with a polynomial Lagrangian and a symmetry-breaking potential. In light of this, we have added a symmetry-breaking potential to cubic galileons, DBI galileons, and conformal galileons. For the first two cases, we found that the results are insensitive to the existence of the galileon terms. This is simply the statement that, when looking for a domain wall solution, the problem is effectively one-dimensional and the higher derivative galileon terms are total derivatives. For the case of the cubic conformal galileon, we found that a domain wall solution does not exist. We have also considered the (A)dS galileons which naturally posses potential terms satisfying their generalized galileon symmetries. In these case, the construction faced several problems related to bubble collapse and the field becoming a ghost.

Finally, we have analyzed the case of solitons moving at the speed of light. We found that any single scalar field in flat space with no potential has a solution of the form $\phi(u)$ or $\phi(v)$, where $u,\, v$ are light-cone coordinates. This means that moving solitons exist for both DBI and conformal galileons. The former being stable and the latter unstable depending on the background soliton. We have traced the instability of the moving conformal solitons to the presence of a term in the Lagrangian that breaks the shift-symmetry. 

\begin{acknowledgments}
We thank Austin Joyce, Garrett Goon, Kurt Hinterbichler, Ken Olum and Alex Vilenkin for useful discussions. The work of M.C. and M.T was supported in part by NASA ATP grant NNX11AI95G. The work of A.M. was supported by NSF grant: PHY-1213888. The work of A.S. was supported by funds provided to the Center for Particle Cosmology by the University of Pennsylvania. M.T. was also supported in part by US Department of Energy (HEP) Award DE-SC0013528.
\end{acknowledgments}

\appendix
	
\section{Generalized galileon Lagrangians}\label{app:gglag}

In this Appendix we present the Lagrangians for the specific generalized galileon theories studied in this paper. As in many treatments of galileons (e.g. \rcite{Goon:2011qf}), we will find it convenient to establish some standard time-saving notation. We will denote by $\Pi$ the matrix of second derivatives of $\pi$, $\Pi_{\mu\nu}\equiv \nabla_\mu\nabla_\nu\pi$. Square brackets denote traces, e.g., $[\Pi] = \Pi^\mu_\mu = \Box\pi$ and $[\Pi^2]=\Pi_{\mu\nu}\Pi^{\mu\nu} = \nabla_\mu\nabla_\nu\pi\nabla^\mu\nabla^\nu\pi$. Finally, we will use the notation $[\pi^n]$ for contractions of $\Pi$ and $\nabla\pi$, defining $[\pi^n]\equiv \nabla\pi \cdot \Pi^{n-2} \cdot \nabla\pi$, e.g., $[\pi^2] = \nabla_\mu\pi \nabla^\mu\pi$ and $[\pi^3] = \nabla_\mu\pi \nabla_\nu\pi \nabla^\mu\nabla^\nu\pi$. Note that for the dS and AdS galileons, $\nabla$ refers to the dS or AdS covariant derivative, and indices are raised and lowered with the dS or AdS metric. 

\subsection{DBI galileons}

We begin with the DBI galileons, which, as mentioned in \cref{sec:gals}, arise from a Minkowski brane probing a Minkowski bulk. The DBI galileon Lagrangians are
\begin{align}\label{eq:DBIgals}
\lag_1 &= \pi,\nonumber \\
\lag_2 &=- \sqrt{1+(\partial\pi)^2}, \nonumber \\
\mathcal{L}_{3}&=-\left [\Pi\right ]+\gamma^{2}\left [\pi^3\right ], \nonumber \\
\mathcal{L}_{4}& =-\gamma \left (\left [\Pi \right ]^{2} -\left [\Pi^{2}\right ]\right )-2\gamma^{3}\left (\left [\pi^{4}\right ]-\left [\Pi\right ]\left [\pi^3\right ]\right ),\nonumber \\
\mathcal{L}_{5}& =-\gamma^{2}\left (\left [\Pi\right ]^{3}+2\left [\Pi^{3}\right ]-3\left [\Pi\right ]\left [\Pi^{2}\right ]\right ) \nonumber \\
&\hphantom{{}=}-\gamma^{4}\left (6\left [\Pi\right ]\left [\pi^{4}\right ]-6\left [\pi^{5}\right ]-3\left (\left [\Pi\right ]^{2}-\left [\Pi^{2}\right ]\right )\left [\pi^3\right ]\right ),
\end{align}
where $\gamma\equiv(1+(\partial\pi)^2)^{-1/2}$.

\subsection{Covariant galileons}

The galileons introduced in \cref{sec:gals} live in Minkowski space. To investigate their dynamics, one might consider simply promoting $\partial\to\nabla$. This is fine for $\lag_2$ and $\lag_3$, but leads to higher derivatives in $\lag_4$ and $\lag_5$. These higher derivatives can be removed with the addition of non-minimal curvature couplings \cite{Deffayet:2009wt}, leading to the covariant galileons,
\begin{align}\label{eq:covgals}
\lag_1 &= \pi, \nonumber \\
\lag_2 &= -\frac{1}{2}(\nabla\pi)^2, \nonumber \\
\lag_3 &= -\frac{1}{2}(\nabla\pi)^2\Box\pi, \nonumber \\
\lag_4 &= -\frac{1}{2}(\nabla\pi)^2\left[(\Box\pi)^2 - \nabla_\mu\nabla_\nu\pi\nabla^\mu\nabla^\nu\pi - \frac R4(\nabla\pi)^2\right], \nonumber \\
\lag_5 &= -\frac{1}{2}(\nabla\pi)^2\left[(\Box\pi)^3 - 3\Box\pi\nabla_\mu\nabla_\nu\pi\nabla^\mu\nabla^\nu\pi\right. \nonumber \nonumber \\
&\left.\hphantom{{}=(\nabla\pi)^2\big[}+ 2\nabla_\mu\nabla_\nu\pi\nabla^\nu\nabla_\alpha\pi\nabla^\alpha\nabla^\mu\pi - 6G^\mu{}_\nu\nabla_\mu\pi\nabla_\alpha\pi\nabla^\alpha\nabla^\nu\pi\right],
\end{align}
where $R$ is the Ricci scalar and $G_{\mu\nu}$ is the Einstein tensor. These have second-order equations of motion for both $\pi$ and $g_\mn$.

\subsection{Conformal galileons}

The conformal galileons \cite{Nicolis:2008in,deRham:2010eu,Khoury:2011da} arise from a derivative expansion of the conformal DBI galileons, which come from taking a flat brane in an AdS bulk. Their Lagrangians are given by
\begin{align}\label{eq:confgals}
\lag_1&=-\frac{1}{4}e^{-4\pi},\nonumber \\
\lag_2&=-\frac{1}{2} e^{-2\pi}(\partial\pi)^2,\nonumber \\
\lag_3&=-\frac{1}{2}\left[(\partial \pi)^2 \Box \pi - \frac12(\partial \pi)^4\right],\nonumber \\
\lag_4&=-\frac{1}{2}e^{2\pi}(\partial \pi)^2\left[[\Pi]^2-[\Pi^2]+\frac25\left((\partial \pi)^2 \Box \pi-[\pi^3]\right)+\frac{3}{10}(\partial \pi)^4\right],\nonumber \\
\lag_5&=-\frac{1}{2}e^{4\pi} (\partial \pi)^2 \bigg[[\pi]^3-3[\pi][\pi^2]+2[\pi^3]+3(\partial \pi)^2([\pi]^2-[\pi^2])\nonumber \\
&\hphantom{{}=e^{4\pi} (\partial \pi)^2 \bigg[}+\frac{30}{7}(\partial \pi)^2((\partial \pi)^2[\pi]-[\pi^3])-\frac{3}{28}(\partial \pi)^6\bigg].
\end{align}
Note that these live on flat space.

\subsection{(A)dS galileons}

The de Sitter and anti-de Sitter galileons can be constructed by taking the small-field limit of the dS and AdS DBI galileons, which arise from a dS brane (in any bulk) and an AdS brane (in an AdS bulk), respectively. Both the dS and the AdS galileons, as well as the standard flat-space galileons, can be combined in the single set of Lagrangians
\begin{align}\label{eq:AdSgals}
\lag_1&=\sqrt{-g}\pi,  \nonumber \\
\lag_2&=-\frac{1}{2}\sqrt{-g} \left((\partial\pi)^2-\frac R3 \pi^2\right),\nonumber \\
\lag_3&=-\frac{1}{2}\sqrt{-g}\left((\partial\pi)^2\Box\pi+\frac R2 \pi(\partial\pi)^2-\frac{R^2}{18}\pi^3\right),\nonumber \\
\lag_4&=-\frac{1}{2}\sqrt{-g}\left[(\partial\pi)^2\left([\Pi]^2-[\Pi^2]+\frac{R}{24}(\partial\pi)^2+\frac{R}{2}\pi\Box\pi+\frac{R^2}{8}\pi^2\right)-\frac{R^3}{144}\pi^4\right], \nonumber \\
\lag_5&=-\frac{1}{2}\sqrt{-g}\left[\left((\partial\pi)^2+\frac{R}{60} \pi^2\right)\left([\Pi]^3-3[\Pi][\Pi^2]+2[\Pi^3]\right)\right.  \nonumber\\ 
&\left.\hphantom{{}=\sqrt{-g}\bigg[}+\frac{2R}{5} \pi(\partial \pi)^2\left([\Pi]^2-[\Pi^2]+\frac{3R}{16}\pi\Box\pi +\frac{5R^2}{144} \pi^2\right)-\frac{R^4}{2160} \pi^5\right],
\end{align}
where $R$ is the scalar curvature of the background space. This links the standard and (A)dS galileons through $R$, which vanishes in flat space and is positive (negative) for (anti)-de Sitter. We remind the reader that the metric used to raise and lower indices, to define covariant derivatives, and which appears in $\sqrt{-g}$ is the metric of (four-dimensional) flat space or (anti)-de Sitter space, depending on the value of $R$.

\bibliography{bibliography}

\end{document}